\documentclass{desyproc}
\usepackage{graphicx}
\begin{document}
\title{The Belle-II Pixel Vertex Detector at the \\ SuperKEKB Flavor Factory}

\author{{\slshape F. Simon$^{1,2}$, K. Ackermann$^1$,  L. Andricek$^3$, C. Heller$^1$, C. Kiesling$^1$, A. Moll$^{1,2}$, H-G. Moser$^3$,  J. Ninkovic$^3$, K. Prothmann$^{1,2}$, B. Reisert$^1$, R. Richter$^3$, M. Ritter$^1$, S. Rummel$^3$, A Wassatsch$^3$ for the DEPFET Collaboration}\\[1ex]
$^1$Max-Planck-Institut f\"ur Physik, M\"unchen, Germany\\
$^2$Excellence Cluster ÔUniverseÕ, Garching, Germany\\
$^3$Max-Planck-Institut f\"ur Physik, Halbleiterlabor, M\"unchen, Germany}
\contribID{PosterID 50}

\confID{800}  
\desyproc{DESY-PROC-2009-xx}
\acronym{LP09} 
\doi  

\maketitle

\begin{abstract}
An upgraded  asymmetric $e^+e^-$ flavor factory, SuperKEKB, is planned at KEK. It will deliver a luminosity of $8\, \times \,  10^{35} \text{cm}^{-2}\text{s}^{-1}$, allowing precision measurements in the flavor sector which can probe new physics well beyond the scales accessible to direct observation. The increased luminosity also requires upgrades of the Belle detector. Of critical importance here is a new silicon pixel vertex tracker, which will significantly improve the decay vertex resolution. This new detector will consist of two detector layers close to the interaction point, using DEPFET pixel sensors with 50 $\mu$m thick silicon in the active area. 
\end{abstract}

\section{Introduction}

The study of $B$ meson systems at the asymmetric energy $e^+e^-$ flavor factories KEKB and PEP-II, running predominantly at the $\Upsilon(4s)$ resonance,  has helped to firmly establish the CKM picture of quark mixing and CP violation in the Standard Model. In addition, new mesonic states involving heavy quarks have been discovered and have been investigated in detail. Despite the success of the current experiments, which have collected 553 fb$^{-1}$ in the case of BaBar and more than 1 ab$^{-1}$ in the case of Belle, many questions yet remain to be answered. With a significant increase in statistical precision, flavor factories have the potential to probe new physics through the deviation of observables from the Standard Model predictions, most spectacularly illustrated by the potential observation of a non-closing unitarity triangle. Such precision measurements thus provide sensitivity to high mass scales, complementary to the direct searches at high energy colliders \cite{Browder:2008em, Browder:2007gg}.

To provide the required increase in statistics, a next generation machine, the super flavor factory SuperKEKB, is planned at KEK \cite{Hashimoto:2004sm}. This upgrade of the existing KEKB collider will provide an instantaneous luminosity of up to $8\, \times \,  10^{35} \text{cm}^{-2}\text{s}^{-1}$, enabling the collection of an integrated luminosity of 50 ab$^{-1}$ in less than a decade of operation. This increase in luminosity by a factor of 40 over the present world record, held by KEKB, will be achieved by extreme focusing of the beams in the interaction region, so-called nanobeams, and by a moderate increase of the beam currents. The increased interaction rates and larger background levels require significant upgrades of the existing Belle detector. For the Belle-II experiment, all major detector subsystems will be replaced or upgraded \cite{Hashimoto:2004sm, Adachi:2008da, Dolezal:2009wq}. 

\section{The pixel vertex tracker PXD}

\begin{figure}
\centerline{\includegraphics[width=0.85\textwidth]{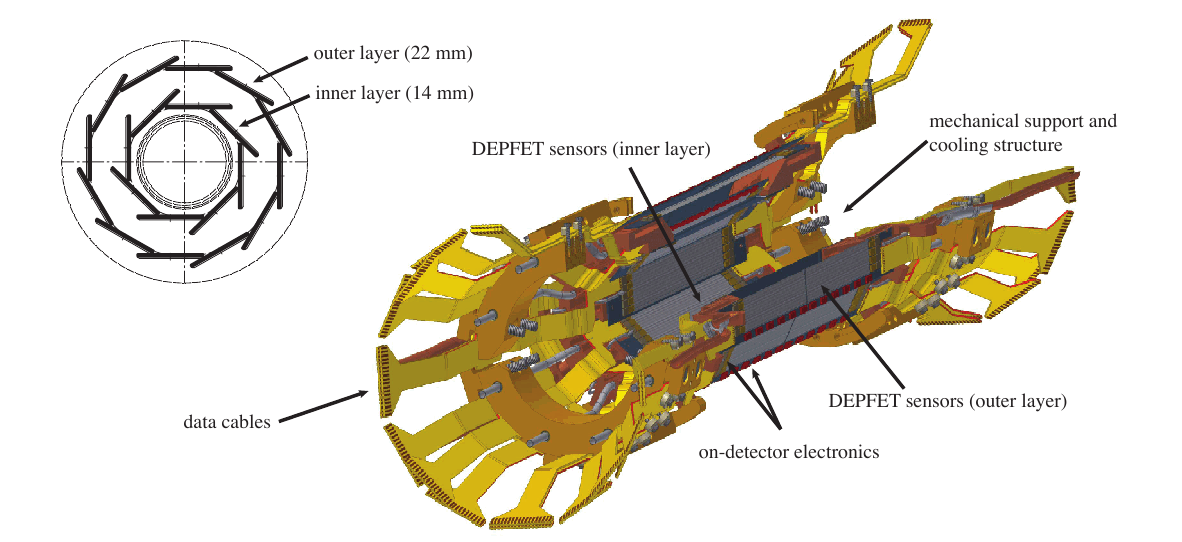}}
\caption{Illustration of the Belle-II PXD, showing the arrangement of the modules of the inner and outer layers (inset upper left), and a the schematic of the overall mechanical design, including support and cooling structures and data cables.}\label{fig:PXD}
\end{figure}

Improved secondary vertex resolution will be provided by a new silicon pixel vertex tracker, the PXD. Excellent vertexing is crucial for  time dependent CP violation studies, where the time difference between the decays of two $B$ mesons is measured by the spatial separation of their decay vertices. This exploits the forward boost of the meson rest frame, provided by the energy asymmetry of the two colliding beams. In the case of $B^0$ pairs, the decay time differences are typically of the order of 1.5 ps, corresponding to flight distances of around 200 $\mu$m.

\begin{wrapfigure}{r}{0.5\textwidth}
\centerline{\includegraphics[width=0.4\textwidth]{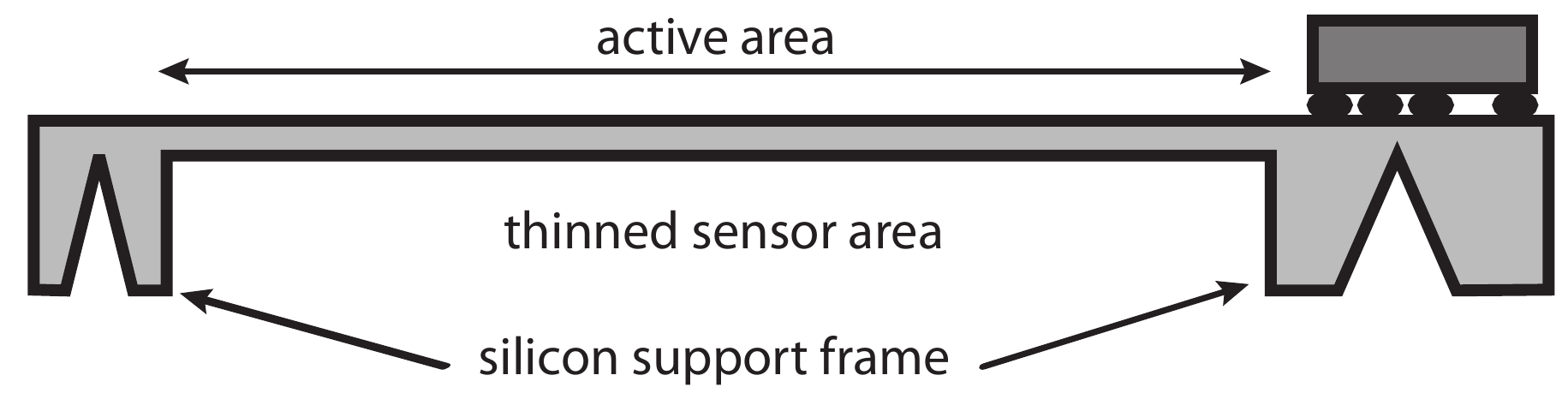}}
\caption{Illustrative cross section through a PXD module. The silicon in the active area is thinned to 50 $\mu$m, 450 $\mu$m thick silicon rims are left on the module sides for stability.}\label{fig:CrossSection}
\end{wrapfigure}

The PXD uses DEPFET \cite{Kemmer:1986vh} active pixel sensor technology, developed at the semiconductor laboratory of the MPI for Physics in Munich. It consists of two layers, at a radius of 14 mm and 22 mm, respectively, arranged around the thin straight section of the Beryllium beam pipe in the interaction region of the Belle-II detector, as illustrated in the inset in the upper right corner of Figure \ref{fig:PXD}. The inner layer uses 8, the outer layer 12 individual detector modules, with an active length of 120 mm for the outer layer, sufficient to cover the full acceptance of the Belle-II detector from 17$^\circ$ to 150$^\circ$, relative to the high energy beam axis. The modules of the inner layer are designed as monolithic silicon structures, while the outer layer modules consist of two joined sensors, required by the size of the 6'' wafers used for production. To minimize the amount of material, crucial for the precise reconstruction of low momentum tracks, the modules themselves are thinned down to a thickness of 50 $\mu$m in the active area, with 450 $\mu$m thick silicon rims to provide structural stability, as illustrated in Figure \ref{fig:CrossSection}. Thus, no additional mechanical support within the detector acceptance is necessary. 

Both inner and outer layer sensors will have 1600 by 250 pixels, which are continuously read out in a rolling shutter mode, with a readout time of 20 $\mu$s. The pixels on the inner modules have a size of roughly $50\,\times\, 50 \, \mu\text{m}^2$, while the outer modules have slightly elongated pixels with a size of  $75\,\times\, 50 \, \mu\text{m}^2$. The read out electronics are located on the module ends, with additional chips for the addressing of individual pixel rows mounted on one side of the stave. The total power consumption is around 7 W per module side, with about 0.5 W dissipated over the active area, and 0.5 W dissipated in the switcher electronics on the stave edges. The required cooling capacity imposes strict constraints on the detector support. The still evolving design is based on support rings on both detector ends made of copper, chosen for the excellent thermal conductivity, which hold the silicon modules. The rings themselves contain cooling channels for active cooling, and through-going air channels to provide cooling for the sensor surfaces with a cold air flow. Several options for the cooling of the support rings are under investigation, including water cooling and evaporative cooling schemes using C$_3$F$_8$ or CO$_2$. The current mechanical design of the PXD, including support structures and data cables, is illustrated in Figure \ref{fig:PXD}.

The DEPFET sensors used in the PXD combine particle detection and amplification of the signal by embedding a field effect transistor into fully depleted silicon. The low noise due to the low capacity of the pixels and the internal amplification provides very high signal to noise ratios. With non-thinned 450 $\mu$m thick sensors using an ILC-like pixel geometry of  $24\,\times\, 24 \, \mu\text{m}^2$ pixels, a signal to noise ratio of  around 130 and a spatial resolution of better than 2 $\mu$m has been achieved in beam tests \cite{Marinas:2009su}. For the thinned sensors, signal to noise ratios in excess of 20 are expected.

Due to the high number of pixels and the continuous readout in a background dominated environment, the zero suppressed data rate of the PXD is expected to be on the order of 20 GB/s. A data reduction scheme using fast online tracking in the silicon strip detector to select regions of interest in the PXD for further data processing is being developed to reduce the PXD data output to a manageable level.

\section{Outlook}

A test production of thinned DEPFET sensors with the Belle-II PXD pixel geometry is well advanced, and will undergo thorough testing in the summer of 2010. Prototypes of the mechanical support and cooling structures are under construction. In parallel, the final detector parameters are being determined with detailed single track and physics simulation studies. The assembly and installation of the PXD in the Belle-II detector and the beginning of data taking at the SuperKEKB flavor factory is envisaged for the year 2013.


\begin{footnotesize}


\end{footnotesize}


\end{document}